\documentstyle[psfig]{mn}
\newif\ifAMStwofonts

\ifoldfss 
  \ifCUPmtlplainloaded \else 
    \NewTextAlphabet{textbfit} {cmbxti10} {} 
    \NewTextAlphabet{textbfss} {cmssbx10} {} 
    \NewMathAlphabet{mathbfit} {cmbxti10} {} 
    \NewMathAlphabet{mathbfss} {cmssbx10} {} 
  \fi 
  \ifAMStwofonts 
    \ifCUPmtlplainloaded \else 
      \NewSymbolFont{upmath} {eurm10} 
      \NewSymbolFont{AMSa} {msam10} 
      \NewMathSymbol{\upi}     {0}{upmath}{19} 
      \NewMathSymbol{\umu}     {0}{upmath}{16} 
      \NewMathSymbol{\upartial}{0}{upmath}{40} 
      \NewMathSymbol{\leqslant}{3}{AMSa}{36} 
      \NewMathSymbol{\geqslant}{3}{AMSa}{3E}

    \fi 
  \fi 
\fi 
 
\ifnfssone 
  \newmathalphabet{\mathit} 
  \addtoversion{normal}{\mathit}{cmr}{m}{it} 
  \addtoversion{bold}{\mathit}{cmr}{bx}{it} 
  \newmathalphabet{\mathbfit} 
  \addtoversion{normal}{\mathbfit}{cmr}{bx}{it} 
  \addtoversion{bold}{\mathbfit}{cmr}{bx}{it} 
  \newmathalphabet{\mathbfss} 
  \addtoversion{normal}{\mathbfss}{cmss}{bx}{n} 
  \addtoversion{bold}{\mathbfss}{cmss}{bx}{n} 
  \ifAMStwofonts 
    \ifCUPmtlplainloaded \else 
      \UseAMStwoboldmath 
      \makeatletter 
      \new@mathgroup\upmath@group 
      \define@mathgroup\mv@normal\upmath@group{eur}{m}{n} 
      \define@mathgroup\mv@bold\upmath@group{eur}{b}{n} 
      \edef\UPM{\hexnumber\upmath@group} 
      \new@mathgroup\amsa@group 
      \define@mathgroup\mv@normal\amsa@group{msa}{m}{n} 
      \define@mathgroup\mv@bold\amsa@group{msa}{m}{n} 
      \edef\AMSa{\hexnumber\amsa@group} 
      \makeatother 
      \mathchardef\upi="0\UPM19 
      \mathchardef\umu="0\UPM16 
      \mathchardef\upartial="0\UPM40 
      \mathchardef\leqslant="3\AMSa36 
      \mathchardef\geqslant="3\AMSa3E 
    \fi 
  \fi 
\fi 
 
\ifnfsstwo 
  \DeclareMathAlphabet{\mathbfit}{OT1}{cmr}{bx}{it} 
  \SetMathAlphabet\mathbfit{bold}{OT1}{cmr}{bx}{it} 
  \DeclareMathAlphabet{\mathbfss}{OT1}{cmss}{bx}{n} 
  \SetMathAlphabet\mathbfss{bold}{OT1}{cmss}{bx}{n} 
  \ifAMStwofonts 
    \ifCUPmtlplainloaded \else 
      \DeclareSymbolFont{UPM}{U}{eur}{m}{n} 
      \SetSymbolFont{UPM}{bold}{U}{eur}{b}{n} 
      \DeclareSymbolFont{AMSa}{U}{msa}{m}{n} 
      \DeclareMathSymbol{\upi}{0}{UPM}{"19} 
      \DeclareMathSymbol{\umu}{0}{UPM}{"16} 
      \DeclareMathSymbol{\upartial}{0}{UPM}{"40} 
      \DeclareMathSymbol{\leqslant}{3}{AMSa}{"36} 
      \DeclareMathSymbol{\geqslant}{3}{AMSa}{"3E} 
    \fi 
  \fi 
\fi 
 
\ifCUPmtlplainloaded \else 
  \ifAMStwofonts \else 
    \def\upi{\pi} 
    \def\umu{\mu} 
    \def\upartial{\partial} 
  \fi 
\fi 
 \title[The Deutsch Field Pulsar I]
 {The Deutsch Field Gamma-Ray Pulsar \\ Paper I : The Model Basics}

\author[M.G. Higgins \& R.N. Henriksen]{M.G. Higgins
 \& R.N. Henriksen \\
 Queen's University, Kingston, Ontario, Canada}
\date{Submitted 1996 January 25}
\pagerange{\pageref{firstpage}--\pageref{lastpage}}
\pubyear{1996}

\def\LaTeX{L\kern-.36em\raise.3ex\hbox{a}\kern-.15em
    T\kern-.1667em\lower.7ex\hbox{E}\kern-.125emX}

\begin{document}

\label{firstpage}

\maketitle

\begin{abstract}

A new model for the high-energy emission from pulsars is developed by 
considering charged particle motion in the fields of a spinning, highly
magnetised and conducting sphere in vacuum. A generally applicable approximation
to the particle motion in strong fields is developed and applied to 
the numerical modelling, and the radiation emitted by curvature emission is
summed to generate light curves. The model predicts many of the observed 
features 
of pulsar light curves. This paper outlines the basic properties of the model;
a subsequent paper will discuss the statistical properties of a population of
model pulsars and apply the model to the known gamma-ray pulsars.

\end{abstract}


\section{Introduction}

Much work has been done over the past twenty-five years on the low-frequency
radio emission detected from pulsars, in large part due to the huge volume of
data available. In the last decade, however, attention has been
turning toward the high end of the emission spectrum, as more reliable data 
become available (see, for example, Grenier et al 1993 and Masnou et al 
1994). 

The main problem in developing a theoretical model of the pulsar magnetosphere
is the complexity of finding a self-consistent solution to the set of equations
defining the electromagnetic fields and particle densities and velocities as
functions of time. Theoretical attemps to model the high-energy emission 
generally make assumptions about the global structure of the magnetosphere,
and then restrict their attention to local sites where large electric fields 
parallel to the magnetic field lines accelerate charges. The two
most developed examples are the polar cap model of Daugherty \& Harding (1994)
and the outer gap model created by Cheng, Ho, \& Ruderman (1986) and further
developed by Chiang \& Romani (1994).

We make a set of assumptions about the global magnetospheric 
structure that differs from 
past models: we assume that the charge
density is very small in the outer magnetosphere, and that the pulsar can be 
treated as a spinning, highly
magnetised and conducting sphere in vacuum (the electromagnetic fields around
such a star were derived
by Deutsch 1955). This assumption is justified in our model because we find that
any charges
created in the inner magnetosphere follow paths constrained to stay
close to the star. They are thus unable to populate the outer magnetosphere and
short out the electric fields (the distinction between the inner and outer
magnetosphere is elaborated in section~\ref{sec-particle_motion}).

By so fixing the fields, particle motion can be integrated relatively
easily, and the radiation observed from the model pulsar can be estimated fairly
accurately. With low number densities, inverse Compton 
scattering and photon/photon interactions are unlikely, and synchrotron emission
is not important, so curvature radiation
is the only source of emission. The back reaction on the particles of the
curvature emission is included in the determining the particle energies.

An approximation to the particle motion in strong fields in developed which
is applicable in all realistic pulsar magnetospheres where gamma-ray production
is efficient. The approximation is used to simplify our numerical integration, 
but is general enough that it may find application in other
models of the pulsar magnetosphere.

In this paper (Paper I), the basic characteristics of the model will be
examined. Higgins \& Henriksen (1996)
(hereafter Paper II) will apply the model to the known high-energy
pulsars, consider how the model pulsar ages, and estimate the number of
high-energy pulsars that should be found in gamma-ray surveys.

\section{The Deutsch Fields and the Vacuum Approximation}

\subsection{The Deutsch Fields}

\begin{figure}
\hspace*{\fill}\psfig{file=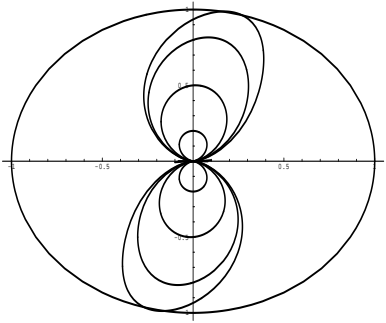,height=3.5in,width=2in}
\hspace*{\fill}\caption[]
{Selected closed field lines for the orthogonal Deutsch field pulsar in the
plane perpendicular to the rotation axis. The circle shows the light cylinder,
and the dimensions are in units of the light cylinder radius $R_{\rm lc}$.}
\label{fig-deutsch_closed}
\end{figure}

\begin{figure}
\hspace*{\fill}\psfig{file=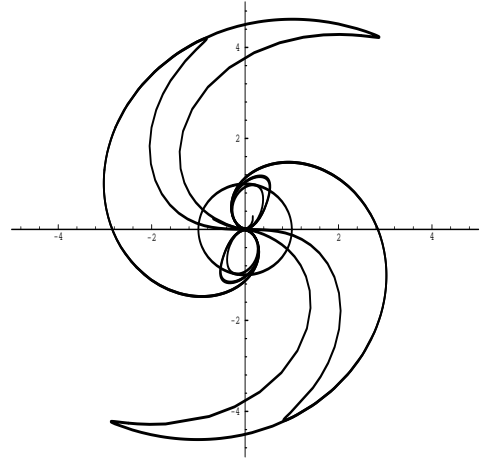,height=5in,width=2.45in}
\hspace*{\fill}\caption[]
{Selected open field lines for the orthogonal Deutsch field pulsar in the
plane perpendicular to the rotation axis. The circle shows the light cylinder,
and the dimensions are in units of the light cylinder radius $R_{\rm lc}$.}
\label{fig-deutsch_open}
\end{figure}

By fixing the electromagnetic fields, the problem of calculating the
radiation pattern changes from a system of partial differential equations to
a system of ordinary differential equations, and becomes much more tractable.

The Deutsch fields look like the fields of a spinning magnetic point dipole
at radii much larger than the stellar radius, with wrapped-up magnetic field
lines outside the light cylinder and large components of electric field
parallel to the magnetic field. Inside the star the electric field is
frozen-in, and this boundary condition at the stellar surface significantly
perturbs the fields near the star. Field lines in the plane perpendicular to 
the rotation axis
for the orthogonal Deutsch fields (when the magnetic dipole moment is 
perpendicular to the rotation axis) near the light cylinder are shown in 
Figure~\ref{fig-deutsch_closed} for selected closed field lines and 
Figure~\ref{fig-deutsch_open} for open field lines.

Four parameters entirely characterise these fields: the angular velocity of the
star $\Omega$, the radius of the star $a$, the magnitude of the effective 
magnetic dipole moment $\mu$, and the inclination $\chi$ of the dipole moment 
with respect to the rotation axis.

An estimate of the electric and magnetic field strength is useful for the
calculation of various quantities; the order of magnitude of the electric and 
magnetic fields near the light cylinder is approximately

\begin{equation}
E \simeq B \simeq \frac{\mu \Omega^3}{c^3}
\end{equation}

The vacuum approximation holds as long as the charge densities and current
densities give negligible perturbations to the fields. This is true when
the charge density is much less than the Goldreich-Julian charge density:

\begin{equation}
\rho_{\rm GJ} \simeq \frac{\vec{\Omega} \cdot \vec{B}}{2 \pi c}
\end{equation}

\noindent where the fields are frozen-in. Here, $\vec{\Omega}$ is the angular
velocity of the star and $\vec{B}$ is the magnetic field at the point in
question. This corresponds approximately to a number density:

\begin{equation}
n_{\rm GJ} \simeq 4 \times 10^{4} {\rm cm}^{-3} (\frac{\Omega}{100 {\rm rads/s}})^4
(\frac{\mu}{10^{30} {\rm G-cm}^3}) (\frac{R_{\rm lc}}{r})^3
\label{eq-GJ}
\end{equation}

\noindent where $R_{\rm lc} = c/\Omega$ is the radius of the light cylinder.
Note the strong variation with the angular
velocity of the star; a relatively small charge density at the
light cylinder of a slow pulsar can alter the fields.

The fields far from the star look like a spherically-propagating electromagnetic
wave with frequency equal to the rotational frequency of the star. This
radiation removes rotational energy from the star, slowing it with time. 
Therefore, the effective magnetic dipole moment of the Deutsch fields can be
estimated from the period and period derivative of the pulsar rotation:

\begin{equation}
\label{eq-mu_est}
\mu \simeq 3.2 \times 10^{37} (P \dot{P})^{1/2} 
\frac{(\frac{I}{10^{45} {\rm g-cm}^2})^{1/2}}{\sin \chi}
\end{equation}

The moment of inertia of neutron stars is unknown by up to a factor of 10
depending on the equation of state; this introduces some uncertainty into the
calculation of the effective magnetic dipole moment.

This result is similar to that of magnetodipole radiation from a point dipole
derived by Ostriker \& Gunn (1969), since the Deutsch fields only differ from
the point dipole fields near the star.

\subsection{Particle-Photon and Photon-Photon Interactions}

This model assumes that the only emission contributing to the light curves
is a result of curvature radiation by the high-energy particles moving along
curved paths in the magnetosphere. This is justified as long as the number
densities are small; how small can be roughly calculated by examining the mean 
free path of photons subject to Compton scattering off electrons and to 
self-scattering.

\subsubsection{Inverse Compton Scattering}

The cross-section for Compton scattering (or inverse Compton scattering) is
approximately equal to the Thompson scattering cross-section for photon
energies in the electron rest frame much less than the electron rest mass.
For photon energies in this frame much greater than the electron rest mass,
the cross-section is approximately 
$\frac{3}{4} \sigma_{\rm T} m c^2/E_{\gamma}$ (Jackson (1975)).
Therefore, the Thompson cross-section will be used as an upper limit to
the cross-section (although this neglects resonant interactions with the
magnetic field, which can increase the cross-section by several orders of
magnitude above the Thompson value for a limited energy range - see, for
example, Daugherty \& Harding 1991).

The optical depth of the magnetosphere to inverse Compton scattering
for a photon is then approximately

\[ \tau_{\rm ICS} \simeq n_{\rm e} \sigma_{\rm T} R_{\rm lc} \]

\noindent where $n_{\rm e}$ is an average electron density, 
$\sigma_{\rm T} = 6.65 \times 10^{-25}$ cm$^2$ is the Thompson scattering
cross-section, and $R_{\rm lc} = c/\Omega$ is the light cylinder radius.

The number density of photons in the magnetosphere can be roughly estimated
from the power output of each electron $P_{\rm e}$, the average photon energy
$E_{\gamma}$, and the characteristic
time the photons are within the magnetosphere $1/\Omega$ as

\begin{equation}
\label{eq-n_gamma}
n_{\gamma} \simeq \frac{P_{\rm e}}{\Omega E_{\gamma}} n_{\rm e}
\end{equation}

To ensure that the number of photons which scatter off electrons is much
less than the number of electrons,

\[ n_{\gamma} \tau_{\rm ICS} \ll n_{e} \]

This puts a limit on the number density of electrons of

\begin{equation}
\label{eq-number}
n_{\rm e} \ll \frac{\Omega^2 E_{\gamma}}{\sigma_{\rm T} c P_{\rm e}}
\end{equation}

For relativistic motion on a path with radius of curvature $\rho$, the power
and average frequency of curvature emission can be written (e.g., Jackson 1975)

\begin{equation}
\label{eq-rad_power}
P_{\rm e} = \frac{2}{3} \frac{e^2 c}{\rho^2} \gamma^4 
\end{equation}

\begin{equation}
E_{\gamma} \simeq \hbar \frac{c}{\rho} \gamma^3 
\end{equation}

\noindent where $\gamma$ is the charge's Lorentz factor and $e$ its charge.

Substituting these into inequality~(\ref{eq-number}), 

\begin{equation}
\label{eq-ne_limit}
n_{\rm e} \ll \frac{\hbar \Omega}{\sigma_{\rm T} e^2 \gamma}
\end{equation}

The Lorentz factor of particles can be estimated from the fields (see
equation~(\ref{eq-Lorentz})) by equating the power output from radiation with the
power input from the electric field. Roughly, this gives

\[ \gamma^4 \simeq \frac{\Omega \mu}{e c} \]

This can be used to define an upper limit to a combination of the stellar
parameters below which this model remains applicable
by setting the Golreich-Julian number density equal to the limit of
equation~(\ref{eq-ne_limit}) (the number densities must always be
much less than the Golreich-Juliann number density). Using the expression
for $\gamma$ above, the condition for neglible scattering is

\begin{equation}
\label{eq-Compton}
\Omega^{\frac{13}{4}} \mu^{\frac{5}{4}} \ll 3.3 \times 10^{49}
\end{equation}

For example, with a magnetic dipole moment of $10^{30}$ G cm$^3$ (giving
a surface magnetic field of approximately $10^{12}$ G), the upper limit on
$\Omega$ is approximately 5000 rad/s, much larger than any standard pulsar
spin rates. Thus, for the population of normal pulsars, the Goldreich-Julian
number density is not sufficient to ensure significant Compton scattering.

\subsubsection{Photon-Photon Scattering}

A similar calculation can be done for the case of photon-photon scattering,
where the cross-section above the pair production limit is again maximally of 
order the Thompson
cross-section. For the spectrum to remain unperturbed as it leaves the
magnetosphere,

\[ \tau_{\gamma-\gamma} \ll 1 \]

\noindent where $\tau_{\gamma-\gamma} = n_{\rm photon} \sigma_{\rm T} 
R_{\rm lc}$.
This gives exactly the same constraint on the electron number density and
on the angular velocity and dipole moment (equation~(\ref{eq-Compton})) as
the inverse Compton scattering criterion above.

While the fraction of photons affected by photon-photon scattering is rather
small, the photons which do interact create charges in the magnetosphere.
This could provide a source for the charges, though a more complex calculation
is required to properly examine the viability of this option.

\subsection{Synchrotron Radiation}

A key approximation of the model (see section~\ref{sec-DFB}) is that charges 
move parallel to the magnetic
field line in a frame where the electric and magnetic fields are parallel.
Therefore, in the framework of the model, there is no contribution at all
from synchrotron radiation. The results of simulating particle motion under
this approximation have been compared to those found in a full-blown simulation
of the particle motion where the approximation is relaxed. The two calculations
agree
remarkably well whenever the fields and energies are large enough to produce
significant amounts of radiation.

\section{Particle Motion and Emission}
\label{sec-motion}

\begin{figure}
\hspace*{\fill}\psfig{file=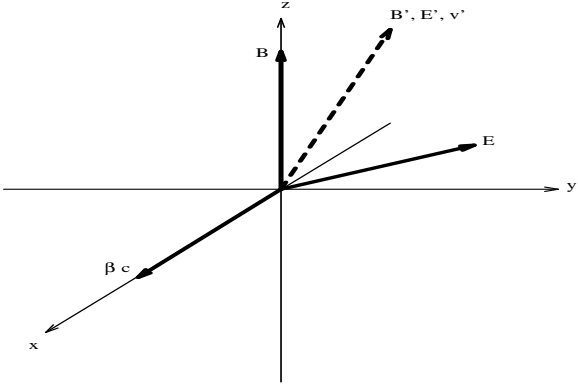,height=2in,width=3in,angle=90}
\hspace*{\fill}\caption[]
{A boost along the $x$-axis shown in the figure moves into a frame where
the electric field $\vec{E'}$ is parallel to the magnetic field $\vec{B'}$. 
In the pulsar
rest frame, the $z$-axis is taken to be parallel to $\vec{B}$, and $\vec{E}$
lies in the $y$-$z$ plane. In the boosted frame, the charges are assumed to
move either parallel or anti-parallel to $\vec{B'}$.}
\label{fig-EB_fig}
\end{figure}

\subsection{The Drift Frame Bead-on-a-Wire Approximation of Particle Motion}
\label{sec-DFB}

The numerical integration of the charged particle motion in the simulations
is simplified enormously under the application of a key assumption in this
work, labelled the drift frame bead-on-a-wire (DFB) approximation. We assume 
that, 
in a frame where the electric field is parallel to the
magnetic field, the charge has a velocity directed parallel or antiparallel
to $\vec{B}$ with magnitude approximately
equal to $c$; the velocity in the pulsar rest frame at any point can then be
determined by boosting back to the original frame of reference.

Consider a point in the magnetosphere with electric field $\vec{E}$ and magnetic
field $\vec{B}$. Define the $z$-axis to be parallel to $\vec{B}$, and 
$\vec{E}$ to have 
components only in the $z$ and $y$ directions. There is a boost of
$v = c \beta \hat{x}$ which defines a frame of reference where $\vec{E}'$ is
parallel to $\vec{B}'$ as long as the electric field is not perpendicular to
the magnetic field {\em and} equal in magnitude (primed quantities are those
in the boosted frame). This is shown in Figure~\ref{fig-EB_fig}; the boost
can be shown to be

\begin{equation}
\beta = \frac{B^2 + E^2 - \sqrt{(B^2 + E^2)^2 - 4 B^2 E_y^2}}{2 B E_y}  
\end{equation}

\noindent where all the quantities relate to the pulsar rest frame. This is
valid for $E > B$ or $E < B$, and even $E = B$ as long as the two are not
perpendicular.

We assume that, in this frame, the velocity of the charge is either parallel
or anti-parallel to $\vec{B}'$ (depending on the sign of $q\vec{E} \cdot \vec{B}$,
where $q$ is the particle's charge)
and that the particle is ultra-relativistic, so its speed is nearly $c$. Boosting
back to the initial frame, the charge's velocity is then

\begin{eqnarray}
\vec{v} & = & c \beta \hat{x} \pm \frac{c}{\gamma} \frac{B_y'}{B'} \hat{y}
\pm \frac{c}{\gamma} \frac{B_z'}{B'} \hat{y} \\
        & = & c \beta \hat{x} \pm \frac{c}{\gamma} 
\frac{\beta E_z}{U} \hat{y}
\pm \frac{c}{\gamma} 
\frac{B - \beta E_y}{U} \hat{z}
\end{eqnarray}

\noindent where $U = \sqrt{B^2 + \beta^2 E^2 - 2 \beta E_y B}$ and 
$\gamma = 1/\sqrt{1 - \beta^2}$. The expression is exact, given
the assumption about the particle motion in the boosted frame of reference.

Given an initial position, the charge's motion is now defined by the solution to
a set of three first-order ordinary differential equations
$\vec{v} = d\vec{r}/dt$, which can be solved 
numerically
with relative ease. The particle's energy can be found by integrating the 
power input due to the electric field and the power output due to the radiation
loss. The simulations of particle motion under this approximation have been
compared to those when the DFB approximation is relaxed in a more sophisticated
(and time-consuming) simulation and are found to match
very closely for all the known gamma-ray pulsars (the constraint on the
applicability of the DFB approximation is considered below).

It is generally found that, since the fields are so large, the radiation 
reaction
force acts like a thermostat, balancing the energy input from the electric
force with the energy output due to radiation. This gives
a convenient description of the particle's energy at any point,

\begin{equation}
\label{eq-Lorentz}
\gamma^4 = \frac{3}{2} \frac{q \vec{E} \cdot \vec{v} \rho^2}{e^2 c}
\end{equation}

\noindent where the radiative power output was taken as 
equation~(\ref{eq-rad_power}).

The whole approximation is only valid where the timescale to radiate away the
momentum perpendicular to the field 
direction is much less than the timescale $1/\Omega$ over which the fields vary 
significantly; this reduces to another constraint on $\Omega$ and $\mu$:

\begin{equation}
\Omega^{\frac{21}{4}} \mu^{\frac{9}{4}} \gg 5 \times 10^{71}
\end{equation}

For a typical $\mu = 10^{30}$ G cm$^3$, this gives $\Omega \gg 6$ rad/s, 
which is true for all the observed gamma-ray pulsars.

\subsection{Types of Particle Motion and Justification of the Vacuum Assumption}
\label{sec-particle_motion}

Three different types of particle motion were observed in the simulations.
Some charges follow open paths and escape to infinity, radiating quite
powerfully inside the magnetosphere; these are labelled ``outgoing'' paths.
The second and third types of path are closed within the inner magnetosphere.
Charges on the second type of path follow a course which directs them into
the star; on the third variety, the charge paths attempt to cross a surface where
$\vec{E} \cdot \vec{B}$ goes to zero. Particles encountering such a
``null surface'' lose their energy almost immediately since the radiation power 
output is
no longer balanced by an energy input from the electric field. Any charge
encountering a null surface is forced to remain on it; the electric
field switches sign on either side of the surface, effectively trapping the
particle. They can travel out of the magnetosphere on these surfaces
at relatively low energies without contributing to the observed radiation, but
perhaps establishing the required charge balance for the steady-state 
magnetosphere. These surfaces then are not so much ``gaps'' as ``conduits'' for
low-energy plasma. 

Which type of path a charge will follow depends on its starting position.
The fraction of paths which are outgoing drops to zero at a radius of 
approximately $0.1 R_{\rm lc}$; {\em all} the charges inside this radius are 
bound to the inner magnetosphere, except those on the surface with
$\vec{E} \cdot \vec{B} = 0$, which can escape the magnetosphere but at
very low energies. This defines the 
boundary between the inner and
outer magnetosphere: inside the inner magnetosphere, all particle paths are
bound.

This constitutes our justification of the evacuated nature of the magnetosphere;
charges created (presumably due to $\gamma-B$ pair creation near the star) will
not be able to populate the outer magnetosphere, since they are locked into
bound orbits. We take the bound charges into account by increasing the effective
stellar radius to $0.1 R_{\rm lc}$. In doing so, the charges
remain bound, even though the field structure changes to become ``frozen-in''.

When charges are started relatively far inside the magnetosphere, but outside
$0.1 R_{\rm lc}$, most of the paths are still bound, and the outgoing paths
extend from a small region near one of the poles (which one depends
on the sign of the charge of the particle).

\section{Outline of the Model}

With this justification of the vacuum assumption for the magnetosphere, the
basic model can be defined as a rough approximation to a global self-consistent
solution to the pulsar problem. The outer magnetosphere ($r > 0.1 R_{\rm lc}$)
is the region of particle acceleration, where charges can be ejected from
the magnetosphere. The only radiation which is important is curvature radiation,
for the reasons discussed earlier. 

Particles radiate gamma-ray photons, some toward the star, where they create
electron-positron pairs on the powerful magnetic fields in the inner 
magnetosphere. These charges are all
bound on closed paths to the inner magnetosphere, and cannot populate the
outer magnetosphere to short out the large electric fields. 
A charge balance is established by having charges being created near the star and
flowing out with low energies 
along the null surfaces. The effect of the outflowing charges on the fields
is neglected.

Since the paths of the charges are all bound for radii $r < 0.1 R_{\rm lc}$,
charges were started on a sphere of radius $0.2 R_{\rm lc}$ (called the
``starting sphere'') at intervals of
$\pi/16$ radians in the toroidal and poloidal angles to fully sample the range
of initial starting positions. The starting energies were unimportant, since
the charges immediately move to balance radiation power out with electric field
power in. The source for these charges is left undetermined.

The radiation from these trajectories was calculated and binned to build up 
light curves and spectra for different viewing directions; this is discussed 
in detail
in sections~\ref{sec-LC_gen} and~\ref{sec-spectra}

To simplify the numerical calculation, we took advantage of three symmetries of
the problem. First, there is a symmetry (Symmetry 1) in the fields $\vec{E} 
\rightarrow
-\vec{E}$ and $\vec{B} \rightarrow -\vec{B}$ under a translation $\phi
\rightarrow \phi + \pi$, $\theta \rightarrow \pi - \theta$, where $\theta$ is
the poloidal angle and $\phi$ is the toroidal angle relative to the rotation
axis. Therefore, if a 
charge $q$ started at $\vec{r}_0$ follows the path $\vec{r}(t)$,
the same charge started at $-\vec{r}_0$ will take the
path $-\vec{r}(t)$. The symmetry applies for any orientation of the dipole
moment to the rotation axis.

The second symmetry (Symmetry 2) is between charges of different sign, but only 
applies
when the dipole axis is perpendicular to the rotation axis.
The path of a charge $-q$ started at $(r_0, \theta, \phi)$ is equal to the
path of a charge $q$ started at $(r_0, \theta, \phi + \pi)$ rotated through
180 degrees toroidally.

The third symmetry (Symmetry 3) is a result of another property of the fields,
that $\vec{E}(r, \theta, \phi, t)$ equals 
$\vec{E}(r, \theta, \phi - \Omega t, 0)$ rotated through an angle $\Omega t$
in $\phi$. This means that a charge $q$ started at $(r_0, \theta, \phi, t_0)$
will follow the path of a charge $q$ started at $(r_0, \theta,
\phi - \Omega t_0, 0)$
rotated through $\Omega t_0$ toroidally. This symmetry holds for any orientation
of the dipole moment.

Symmetry 1 means than runs need be done only for particle starting points in
the top half of the magnetosphere; paths in the bottom half 
are
reflections of those in the top half. A consequence of
Symmetry 3 is that charges started at a time $t_0$ not equal to zero
can be found from the paths started at $t_0 = 0$ by a rotation azimuthally.
Symmetry 2 means that positron paths can be inferred from electron paths,
but only in the orthogonal Deutsch fields.

The combination of Symmetry 2 and Symmetry 3 has an important consequence for 
the inferred light
curves when the dipole moment is perpendicular to the rotation axis:
positrons will radiate in exactly the same manner as
electrons, but a half period later. If the electron contribution is a single
peak, the positrons will contribute a second peak, exactly 0.5 later in phase,
if the distribution of positrons is equal to that of electrons in the
magnetosphere. This does not hold for non-orthogonal systems, as Symmetry 2
is broken.

\begin{figure*}
\hspace*{\fill}\psfig{file=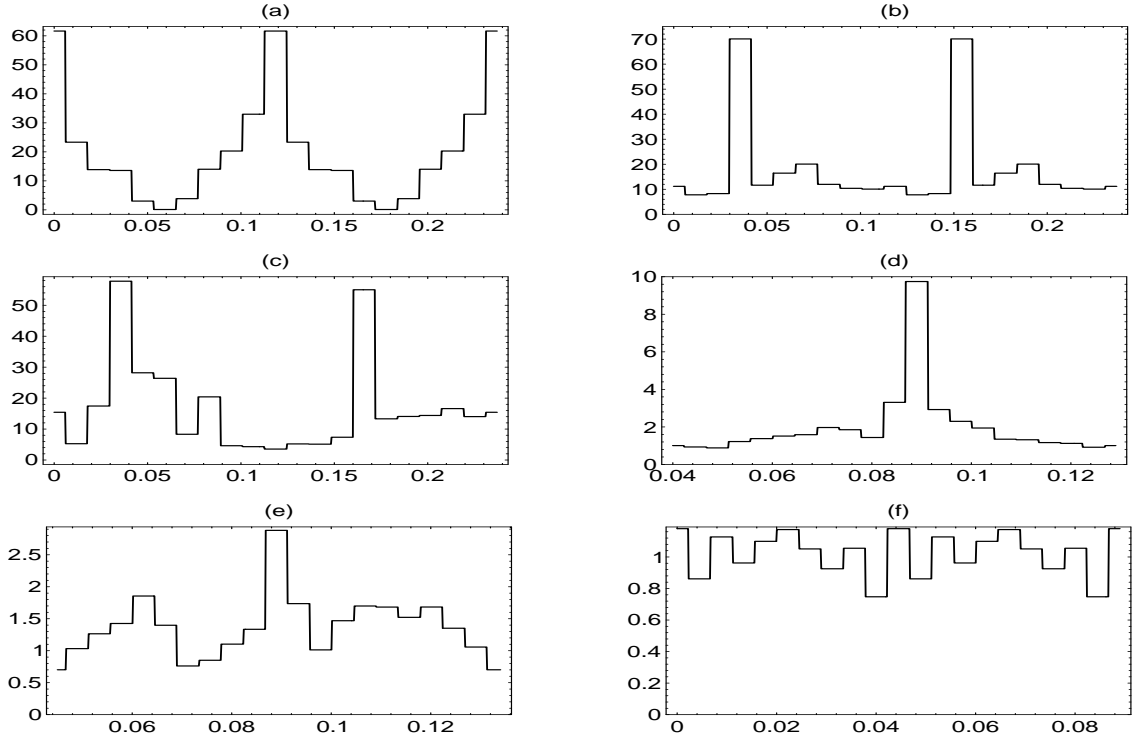,height=5.5in,width=6.5in}
\hspace*{\fill}\caption[]
{Representative light curves produced by the model for different viewing
directions, spin rates, field strengths, and dipole inclinations. See the
text for a complete discussion. Time is plotted along the horizontal axis,
and arbitrary intensity along the vertical. All plots have radio peaks at
$\tau = 0$ and/or $\tau = \pi/\Omega$, presuming the radio emission to be 
generated close to the star
and near the pole. Plot (d) is shown offset in time to show the single
peak more clearly.}
\label{fig-rep_curve}
\end{figure*}

\section{Light Curve Generation}
\label{sec-LC_gen}

With paths of all the charges known, a pattern of radiative emission for
different retarded times $\tau$ ($\tau = t - \hat{n}_{\rm obs} \cdot \vec{r}/c$,
with $\hat{n}_{\rm obs}$ a unit vector pointing toward the observer)
could be generated. 

A numerical routine stepped along the paths for
each of the different starting angles, and for 100 steps in starting angle
$t_0$. The emission from paths in the bottom half of the magnetosphere (and
for positron paths when considering orthogonal fields) was included by 
inferring their paths from the three symmetries discussed earlier.

Since the emission generated is only for a single electron started at
$(\theta, \phi)$ on the starting sphere and at time $t_0$, the power must
be scaled by a distribution function $f_{\rm e,p}$ of charges over the 
starting sphere and over starting time, where $f_{\rm e, p} d\Omega dt_0$ is the 
number
of charges (either electrons or positrons) flowing through the starting sphere 
between solid angle $\Omega$
and $\Omega + d\Omega$ at times between $t_0$ and $t_0 + dt_0$. This
distribution function can be varied to model different sources; however,
since most emission comes from the outgoing paths, which only originate on
a small region of the starting sphere, changes in $f_{\rm e,p}$ do not much 
affect the
light curves. Throughout this work, $f_{\rm e} = f_{\rm p} = f$ for the light
curve generation; the actual value of $f$ is constrained by the total power 
emitted by the pulsar, and sets the number density at the starting sphere. 

The spectrum of the radiation at one point along a path, for a given particle
Lorentz factor and path radius of curvature, is assumed to be instanteously
that for circular motion; this is discussed in section~\ref{sec-spectra}
The emitted radiation was calculated at several different frequencies to capture 
the spectrum at each time on the light curve.

The emitted power was binned into angular bins of $\pi/16$ radians in azimuthal 
and
polodial viewing angles, bins of $T/20$ in $\tau$ (where $T$ is the pulsar
period), and bins of 0.25 in $\log \omega$, where $\hbar \omega$ is the
photon energy. Any emission which was beamed through the inner magnetosphere
($r < 0.1 R_{\rm lc}$) was not included in the light curves, as it was assumed
to have been absorbed in the denser plasma near the star.

Figure~\ref{fig-rep_curve} shows some representative 1 MeV light curves generated
by the model, which cover many of the variations seen in the observed gamma-ray
pulsar light curves.
Plots (a) and (b) correspond to a Geminga-like pulsar, with
an angular velocity of $\Omega = 26.5$ rad/s and a magnetic dipole moment
of $\mu = 10^{30}$ G cm$^3$ aligned perpendicular to the rotation axis.
The first is for a viewing angle of $\theta_{\rm v} = 84.4^\circ$, and
the second represents radiation into $\theta_{\rm v} = 50.6^\circ$.
These show the characteric symmetry between the first and second halves of
the light curve due to the action of Symmetry 2 in the orthogonal field.
Light curve (c) is for the same Geminga-like pulsar, but with the dipole moment
inclined at 45$^\circ$ to the rotation axis, seen at a viewing angle of 
73$^\circ$. The two main peaks are separated by a phase of 0.45.
(d), (e), and (f) all correspond to emission from a Vela-like
pulsar, with $\Omega = 70.6$ rad/s and $\mu = 3.7 \times 10^{30}$ G cm$^3$.
The dipole moment in (d) and (e) is nearly aligned at only 11.25$^\circ$ to the 
rotation axis; (f) corresponds to orthogonal fields. (d) represents emission 
into a viewing angle of 16.9$^\circ$, and shows a well-defined single peak.
(e) is seen from a viewing angle of 61.9$^\circ$ and shows three quite
well-defined peaks (there is a suggestion of a triply-peaked light curve
for PSR B1706-44 - Thompson et al, 1995). (f) is seen from a viewing angle
of $39.4^\circ$, and shows no
discernable pulsation at frequencies less than 50 GeV. This pulsar would likely 
be missed in a pulsar survey.

All the plots discussed relate to emission beamed into $\phi_{\rm v} = 0$;
one of the poles points toward the observer at $t = 0$ at this angle. The
light curves for $\phi_{\rm v} \neq 0$ are identical except for a phase change.
For all these plots, the radio emission peaks (presumed to come from the poles
near the star) would be seen at $\tau \simeq 0$ and/or $\tau = \pi/\Omega$; 
the gamma-ray peaks are 
significantly displaced in some cases, while in others they align quite closely.

\section{Total Emitted Power}

The total power emitted into all directions by the electrons or positrons at 
any time $t$ is given 
by:

\begin{equation} 
P_{\rm e,p}(t) = \int_{\Omega} \int_{t_0} [ f_{\rm e} 
\hat{P}_{\rm e}
+ f_{\rm p}
\hat{P}_{\rm p} ]
d \Omega d t_0
\end{equation}

\noindent where $\hat{P}_{\rm e, p}(\theta, \phi, t_0; t-t_0)$ is the power 
output of one electron (e)
or positron (p) at time $t-t_0$ along the path, started at $(\theta, \phi)$ on 
the starting sphere at time $t_0$. $f_{\rm e, p}$ are the distribution functions
of charges on the starting sphere, which is not fixed in the model.

\subsection{The Required Number Density}

If the total high-energy power output of a pulsar is known, this fact can be 
used
to estimate the number density of charges at the starting sphere required to
supply that energy. If this number density is larger than the Goldreich-Julian
number density, the fields will be altered and the model will break down.

Taking a constant distribution $f_{\rm e} = f_{\rm p} = f$, the total energy
radiated in one period is given by

\begin{equation}
\label{eq-E}
E = f \int_{\Omega} \int_{t_0} \int_{t} 
  (\hat{P}_{\rm e} + \hat{P}_{\rm p}) dt dt_0 d\Omega
\end{equation}

Since the radiation reaction force fixes the radiated power to be very closely
equal to the power input from the electric field,

\begin{equation}
\hat{P}_{\rm e,p} = e \vec{E} \cdot \vec{v}_{\rm e,p} 
= e c \frac{\mu \Omega^3}{c^3} \tilde{E} \cdot \vec{\beta}_{\rm e,p}
\end{equation}

\noindent where $\tilde{E}$ is a dimensionless electric field which no longer
depends on the values of $\mu$ and $\Omega$, and $\vec{\beta}_{\rm e,p} =
\vec{v}_{\rm e,p}/c$.

Defining $\tau_0 = \Omega t_0$ and $\tau = \Omega t$, the total energy emitted
in a period can now be written

\begin{equation}
E_{\rm tot} = f \frac{e \Omega \mu}{c^2} {\cal E}
\end{equation}

\noindent where ${\cal E}$ is a dimensionless function of the dipole inclination
angle $\chi$ alone:

\begin{equation}
{\cal E} = \int_{\Omega} \int_{\tau_0 = 0}^{2 \pi} \int_{\tau = 0}^{2 \pi}
\tilde{E} \cdot (\vec{\beta}_{\rm e} + \vec{\beta}_{\rm p}) d\Omega d\tau_0
d\tau
\end{equation}

For example, ${\cal E} = 36$ for the orthogonal fields, and 6.8 for 
$\chi = 11.25^\circ$.

Since the total gamma-ray power output from the model must be constant with
time, the total energy emitted in a period is equal to $P_{\rm obs} 
\frac{2 \pi}{\Omega}$, where $P_{\rm obs}$ is the observed gamma-ray power.

The value of $f$ can then be related to the observed power output:

\begin{equation}
f = \frac{2 \pi c^2 P_{\rm obs}}{e \Omega^2 \mu {\cal E}}
\end{equation}

The value of $f$ sets the number density at the starting sphere required to 
produce the observed power output, since $f = 4 \pi c (0.2 R_{\rm lc})^2 n$.
Therefore,

\begin{equation}
n = \frac{25 P_{\rm obs}}{2 e c \mu {\cal E}}
\end{equation}

\begin{table*}
\centering
\begin{minipage}{140mm}
\begin{tabular}{|l|c|c|c|c|c|l|} \hline
Pulsar & $P$ & $\dot{P}$ & $\mu$ & $P_{\rm tot}$ & $n/n_{\rm GJ}$ 
& Reference \\
\hline
Geminga & 0.237 & 11.0 & 1.63 & $1.6 \times 10^{34}$ & $5.6 \times 10^{-3}$ & Mayer-Hasselwander et al 1994 \\
Vela & 0.0893 & 125 & 3.4 & $6.3 \times 10^{33}$ & $1.0 \times 10^{-5}$ & Kanbach et al 1994 \\
Crab & 0.0333 & 421 & 3.8 & $10^{36}$ & $2.6 \times 10^{-5}$ & Ulmer et al 1995 \\
PSR1509-58 & 0.150 & 1540 & 15 & $6.3 \times 10^{35}$ & $4.2 \times 10^{-4}$ & Laurent et al 1994 \\
PSR1706-44 & 0.102 & 93.0 & 3.1 & $6.1 \times 10^{34}$ & $2.1 \times 10^{-4}$ & Thompson et al 1992 \\
PSR1055-52 & 0.197 & 5.8 & 1.1 & $9.4 \times 10^{33}$ & $3.5 \times 10^{-3}$ & Fierro et al 1993 \\
PSRB1951+32 & 0.0395 & 5.85 & 0.49 & $1.3 \times 10^34$ & $4.0 \times 10^{-5}$ & Ramamanamurthy et al 1995 \\
\hline
\end{tabular}
\caption{Number densities at the starting sphere required to give the observed
power outputs for the known gamma-ray pulsars, as a fraction of the 
Goldreich-Julian number density. Periods $P$ are given in seconds, period
derivatives $\dot{P}$ in $10^{-15}$ s/s, dipole moments $\mu$ in units of 
$10^{30}$~G~cm$^3$, and gamma-ray power outputs $P_{\rm tot}$ in ergs/s. All 
known high-energy pulsars have $n/n_{GJ} \ll 1$.}
\end{minipage}
\end{table*}

The number density can be described somewhat more meaningfully as a fraction
of the Goldreich-Julian number density (equation~(\ref{eq-GJ})) at the starting 
sphere (${\cal E}$ is taken to be equal to 36, the value for the orthogonal
fields).

\begin{equation}
\frac{n}{n_{GJ}} \simeq 4.7 \times 10^{-5} {\rm cm}^{-3} 
(\frac{P_{\rm obs}}{10^{34} {\rm ergs/s}})
(\frac{\Omega}{100 {\rm rad/s}})^{-4} (\frac{\mu}{10^{30} {\rm G cm}^3})^{-2}
\end{equation}

This quantity is tabulated in Table~1 for the known gamma-ray pulsars, and is
less than unity, and therefore consistent with the model, in all cases. The
values of $\mu$ were calculated with equation~(\ref{eq-mu_est}) taking 
$\sin \chi = 1$ (and are therefore a lower limit).

\begin{figure}
\hspace*{\fill}\psfig{file=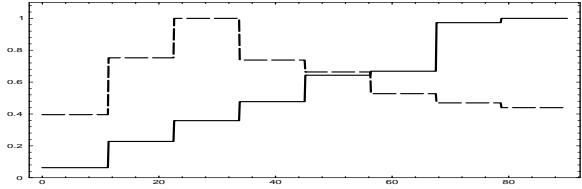,height=2in,width=3in}
\hspace*{\fill}\caption[]
{Distribution of emitted power at 1 MeV with viewing angle for two Vela-like 
pulsars. The solid line shows the power emitted by a pulsar
with orthogonal fields, and the dashed line shows the corresponding curve for
a pulsar with a dipole axis inclined
at 11.25$^\circ$ to the rotation axis. The axes show viewing angle in degrees on 
the horizontal and
emitted power scaled to unity at the maxima on the vertical. }
\label{fig-angular}
\end{figure}

\begin{figure}
\hspace*{\fill}\psfig{file=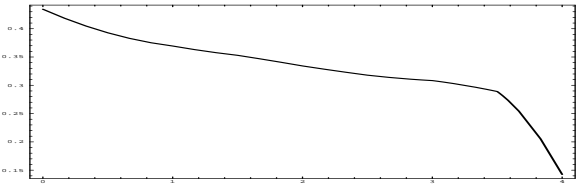,height=2in,width=3in}
\hspace*{\fill}\caption[]
{The variation of $\Omega_{\rm p}$, the fractional solid angle into which
radiation is beamed, is shown for a Geminga-like pulsar with a dipole moment
inclined at 22.5$^\circ$ to the rotation axis. The pulsar becomes significantly
less observable at higher frequencies. Similar results are obtained for
pulsars with different model parameters. The axes show the base-10 logarithm
of the photon energy in MeV along the horizontal, and the value of 
$\Omega_{\rm p}$ along the vertical.}
\label{fig-om_var}
\end{figure}

The most serious problem with this model is its inability to {\em predict} the
total power output; instead, it is only shown that the observed power outputs
require number densities less than the Golreich-Julian number density, and
therefore are consistent with the assumption of vacuum around the star. The
source of the charge on the starting sphere is left unexplained.

\subsection{The Gamma-Ray Efficiency}

The gamma-ray efficiency of a pulsar is defined as the ratio of the power
output in gamma-rays to the total energy loss. For the Deutsch field
pulsar, the rate of rotational energy loss is 

\begin{eqnarray}
P_{\rm rot} & = & \frac{2 \Omega^4 \mu^2 \sin^2 \chi}{3 c^3} \\
            & \simeq & 2.5 \times 10^{-32} \Omega^4 \mu^2
\end{eqnarray}

\noindent and the gamma-ray power emitted in our model is given by

\begin{equation}
P_{\gamma} = \frac{2 {\cal E}}{25} e c \mu n
\end{equation}

Setting ${\cal E} = 36$ (orthogonal fields) and writing $n$ in terms of the
Goldreich-Julian number density $n_{\rm GJ}$, the gamma-ray efficiency is 

\begin{equation}
\eta_{\gamma} = \frac{P_{\gamma}}{P_{\rm rot} + P_{\gamma}} \simeq 
\frac{83 \frac{n}{n_{\rm GJ}}}{1 + 83 \frac{n}{n_{\rm GJ}}}
\end{equation}

Here, the number density has been written as a fraction of the 
Goldreich-Julian number density; the gamma-ray efficiency only depends on this
ratio, not on the pulsar properties directly.

As $n/n_{\rm GJ}$ approaches unity, so does the gamma-ray efficiency. This
is seen in the cases of Geminga and PSR1055-52, where the observations suggest 
that $\eta_{\gamma} \simeq 1$ (Mayer-Hasselwander et al 1994, Fierro et al 1993).

\subsection{The Distribution in Viewing Angle}

The emission is not beamed equally into all directions. Figure~\ref{fig-angular}
shows the emission (averaged over a period) at 1 MeV as a function of the 
viewing angle ($\theta_{\rm v} = 0$ to 90$^\circ$) for two Vela-like pulsars:
one with orthogonal fields (solid
line) and the other with a dipole inclination
of 11.25$^\circ$ (dashed line). The power is scaled to unity at the maximum.

The fraction of the total solid angle into which the power is beamed can be 
quantified in terms of a power-weighted average:

\begin{equation}
\Omega_{\rm p} = \int_{\Omega} \frac{P(\Omega)}{P_{\rm max}} d\Omega
\end{equation}

\noindent where $\Omega$ represents solid angle, not the angular velocity of
the pulsar.

Figure~\ref{fig-om_var} shows $\Omega_{\rm p}$ as a function of frequency for
a Geminga-like pulsar with a dipole moment inclined at 22.5$^\circ$ to the
rotation axis. A general trend to lower $\Omega_{\rm p}$ (and therefore less
observable pulsars) is seen for increasing frequency, with a much faster dropoff
above the break in the spectrum (near several hundred MeV). Similar results are
obtained for pulsars with different model parameters.

\section{Spectra of Emitted Radiation}
\label{sec-spectra}

\begin{figure}
\hspace*{\fill}\psfig{file=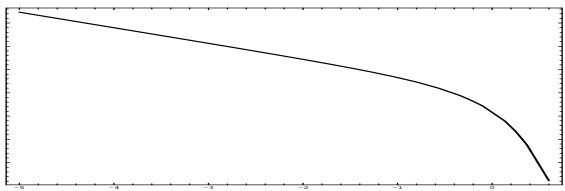,height=2in,width=3in}
\hspace*{\fill}\caption[]
{The curvature spectrum for a single particle in instantaneously circular
motion. Logarithm of photon energy in units of the reference energy
$\hbar \gamma^3 c/\rho$ is shown along the horizontal, and logarithm of 
arbitrary differential photon number is plotted along the vertical.}
\label{fig-single_part_spec}
\end{figure}

Since curvature radiation is the only contribution to the light curves, the 
frequency
distribution of curvature emission by a single particle moving instantaneously
in circular motion must be considered
(see, for example, Jackson (1975)).
The shape of the curve depends on only two parameters - the energy of 
maximum intensity ($E_{\rm m} = 0.14 \hbar \gamma^3 c/\rho$, where $\gamma$ is 
the
charge's Lorentz factor, $c$ is the speed of light, and $\rho$ is the
radius of curvature of the particle's path), and the maximum intensity. 
For energies
much less than $E_{\rm m}$, the number distribution in frequency is proportional
to $E^{-2/3}$; above $E_{\rm m}$, the curve drops off exponentially, and 
essentially disappears for $E > 30 E_{\rm m}$. This function is plotted in 
Figure~\ref{fig-single_part_spec}.

\begin{figure*}
\hspace*{\fill}\psfig{file=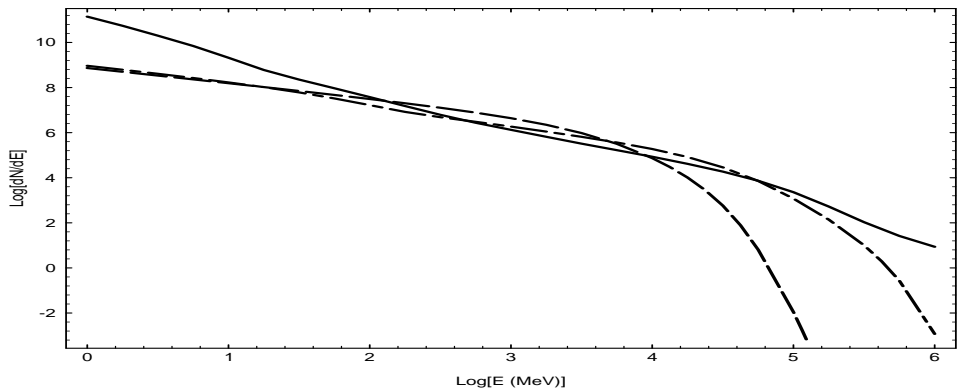,height=4in,width=5in}
\hspace*{\fill}\caption[]
{Representative spectra for three model pulsars, all with $\mu=10^{31}$ G cm$^3$,
$\chi = 90^\circ$, and seen from a viewing angle of $84.3^\circ$. The solid
line corresponds to a pulsar with $\Omega = 150$ rad/s, the dashed-dotted
line to one with $\Omega = 75$ rad/s, and the dashed line to $\Omega = 25$ rad/s.
The faster pulsars show more spectral detail at low frequency.}
\label{fig-rep_spectra}
\end{figure*}

At low frequencies, therefore, each part of the light curve will fall off
in photon number like $E^{-2/3}$, and the shape of the light curve will remain
essentially constant. At higher frequencies, the contributions from particles
radiating with different $E_{\rm m}$'s will result in a light curve which is
variable with frequency. If several different locations in the magnetosphere
make significant contributions to the power, the shape of the spectrum can
deviate significantly from the single-particle spectrum.

Figure~\ref{fig-rep_spectra} shows several different time-averaged spectra from 
model pulsars, all with $\mu = 10^{31}$ G cm$^3$, $\chi = 90^\circ$, and
seen from a viewing angle of $83.4^\circ$. Spectra for pulsars with three
different angular velocities are shown: $\Omega = 25$ rad/s, 75 rad/s, and
150 rad/s. The slow pulsar shows a very hard spectrum with an energy break near 
several GeV, but relatively featureless otherwise. The 75 rad/s pulsar shows
a slight softening of the spectrum near a few hundred MeV, and then returning
to a hard profile until an energy break near 100 GeV. The 150 rad/s pulsar
has a significantly harder spectrum with no break even through the TeV region.
It is interesting to note that the photon flux for all three pulsars is
relatively constant in the 100 MeV range even though the total power outputs
vary considerably.

\begin{figure}
\hspace*{\fill}\psfig{file=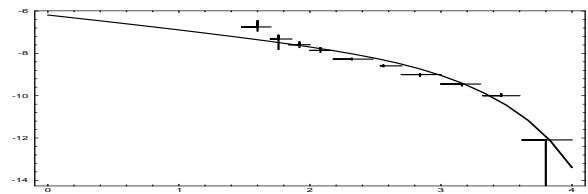,height=2in,width=3in}
\hspace*{\fill}\caption[]
{The model spectrum of a Geminga-like pulsar from 1 MeV to 10 GeV. The data
shown are EGRET data in the range 40 MeV to 6.3 GeV,
from Mayer-Hasselwander et al 1994. The theoretical curve has been shifted
vertically on the plot to best fit the data. The horizontal axis shows the
logarithm of the energy in MeV, and the vertical axis shows the logarithm of
differential photon number in units of photons/s/cm$^2$/MeV.}
\label{fig-spec_plot}
\end{figure}

\begin{figure*}
\hspace*{\fill}\psfig{file=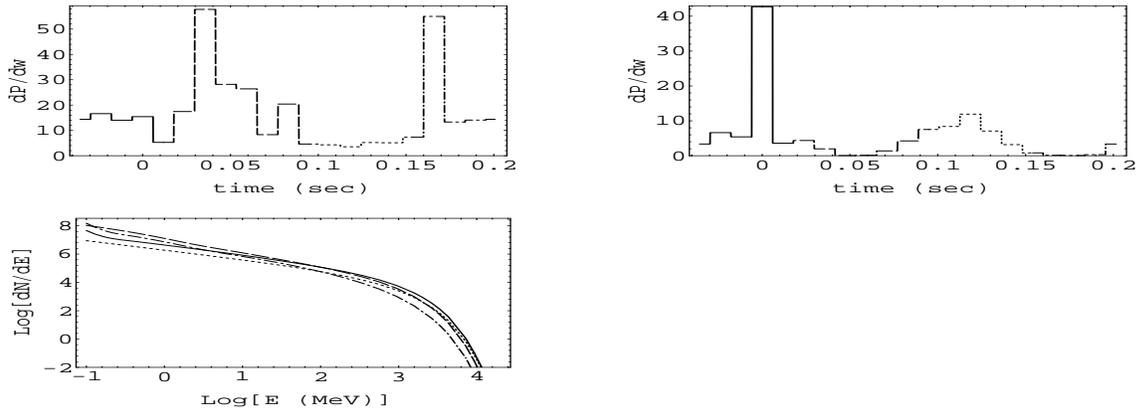,height=5in,width=6.5in}
\hspace*{\fill}\caption[]
{The frequency dependence of the light curve (c) from Figure~4. (a) shows the
light curve at 1 MeV, and (b) shows the emission at 10 GeV. (c) shows the
spectra for four different components in the light curve. The line dashing of
the components in the light curves corresponds to that of the lines in the
spectral plot. The axes show the base-10 logarithm
of energy in MeV along the horizontal and logarithm of arbitrary differential 
photon number along the vertical.}
\label{fig-curve_and_spec}
\end{figure*}

Figure~\ref{fig-spec_plot} shows time-averaged spectra for an orthogonal
field, Geminga-like 
pulsar with a dipole moment of $10^{30}$ G cm$^3$ at a viewing direction of 
$\theta_{\rm v} = 84.4^\circ$. The spectrum
does change slightly with viewing angle, though the major features (location
of the energy break and the shape of the spectrum) are roughly the same. Data
from Mayer-Hasselwander et al 1994 are shown in the interval 40 MeV to 6.3 GeV;
the theoretical curve has been shifted vertically on the plot to best fit the
data.

The different regions of the light curve can have somewhat different spectral
properties. Figure~\ref{fig-curve_and_spec} shows the light curve~(c) from
Figure~\ref{fig-rep_curve} separated into four components. The components
correspond roughly to the peaks at low energy (1 MeV), shown in (a), and
the high-energy (10 GeV) peaks, which fall at phases of zero and 0.5.

The phase of the gamma-ray peak tends to change with energy when the 
energies
are close to the energy break if it is significantly different from the phase of
the radio peak. This occurs because the emission of highest
frequency comes from particles moving nearest to the star, where the fields
are larger. As the energy increases past the break, the peaks tend to move
into phase with the radio peaks, since both are coming from relatively close
to the star near the poles. This alignment of gamma-ray phase with radio phase 
at high frequencies is a provable prediction of the model, given the presumed
location of radio emission. It is difficult to see this effect in the present
observations, since the phase change only happens at very high frequencies
(typically above 10 GeV). However, if the low-energy gamma peaks are not aligned 
with the radio peaks, it does lead to the ``interpulse'' regions
at low energy having harder spectra and higher energy breaks than the low-energy
pulses, as is seen, for example, in the case of the Vela pulsar (Kanbach et al
1994).

Generally the low-energy spectral index of $-2/3$ is harder than observed
in the X-ray region for those pulsars where this 
has been measured. However, this is precisely the region where the model begins 
to break down, and radiation from other areas (such as the inner magnetosphere,
where unperturbed curvature radiation likely is not the dominant contribution 
to the spectrum) becomes important.

\section{Discussion}

A new model of high-energy pulsar emission has been presented, with the
simplifying assumption that the charge densities and current densities are
small enough that the electromagnetic fields can be represented by the
Deutsch fields in the outer magnetosphere.

The assumption that the magnetosphere can remain evacuated is justified
by recognising that charges created near the star follow paths bound to the
inner magnetosphere, and cannot short out the fields farther out. We create
an approximate self-consistent global solution to the electromagnetic fields
and charge motions by extending the ``stellar'' radius (inside which the fields
are frozen-in) to 0.1 of the light cylinder radius; charge
created in the inner magnetosphere can flow out along null surfaces where
the electric field is perpendicular to the magnetic field. The outer
magnetosphere is the location of particle acceleration and high-energy emission.

An important approximation has been developed to chart 
particle motion in fields
where the timescale required to adjust to the field direction is much less
than the timescale to radiate away the charge's momentum perpendicular to the
magnetic field, called the drift frame bead-on-a-wire
(DFB) approximation. This 
reduces the problem of particle
motion to a simple first order, ordinary set of differential equations,
reducing computation time enormously. This approximation could prove
useful for developing global, self-consistent models of the pulsar magnetosphere,
as it gives a relatively simple description of charge paths in terms of the 
electric and magnetic fields.

Several constraints on the applicability of the model can be made, the most
important of which is that the number density required to produce the observed
power output must be much less than the Goldreich-Julian charge density.
All observed gamma-ray pulsars satisfy this constraint, as well as that which
maintains the dominance of curvature radiation to the high-energy spectrum in
the model framework and that which maintains the applicability of the DFB
approximation. 

The light curves generated from the model match the general properties of the
observed light curves
quite well. When the magnetic dipole moment is orthogonal to the rotation
axis, the light curves are mainly double-peaked; the second half of the
light curve is identical to the first. Other inclinations can give single peaks,
non-symmetric double peaks, and triple peaks, as well as practically unpulsed
emission in some cases.

The biggest weakness of the model is its present inability to predict the
actual power output of the pulsars in gamma-rays: it does not assume
any value for the number density at the starting sphere, only that it be
much less than the Goldreich-Julian number density so that the electric fields
are not shorted out. It also says nothing about any radiation produced in
the inner, bound magnetosphere, where particle energies will be somewhat
lower than those in the outer, evacuated magnetosphere. This might provide
significant emission at lower energies.

Our model compares quite favourably with the other popular models for pulsar
gamma-ray emission. The spectra predicted by our model match the observations
in some cases better than those generated by the outer gap model of Chiang \& 
Romani (1994),
and as well as those presented for the extended polar cap model of 
Daugherty \& Harding (1995). Our model also has the potential to generate
more complex light curves than either the outer gap or polar cap models, which
can only produce single or double peaks. However, this is something of a 
double-edged sword, as a random sample of light curves would be somewhat
less regular than is observed.

This work has introduced the general Deutsch field gamma-ray pulsar model;
applications of the model to the known gamma-ray pulsars are developed in
Paper II, which also considers several statistical properties of a population
of these pulsars.

\label{lastpage}

\end{document}